\documentclass[fleqn,usenatbib]{mnras}

\usepackage{newtxtext,newtxmath}
\usepackage{rotating}

\usepackage[T1]{fontenc}

\DeclareRobustCommand{\VAN}[3]{#2}
\let\VANthebibliography\thebibliography
\def\thebibliography{\DeclareRobustCommand{\VAN}[3]{##3}\VANthebibliography}

\usepackage{graphicx}	
\usepackage{amsmath}	
\usepackage{bm}

\title[SNe~Ia progenitor age and cosmic acceleration]
{Pantheon+ supernovae corrected for progenitor age indicate the universe is decelerating}

\author[Sah et al.]{
Animesh Sah,$^{1}$ 
Mohamed Rameez,$^{1}$
Subir Sarkar~$^{2}$
\\
$^{1}$Department of High Energy Physics, Tata Institute of Fundamental Research,Homi Bhabha Road, Mumbai,400005,India\\
$^{2}$Rudolf Peierls Centre for Theoretical Physics, University of Oxford, Parks Road, Oxford, OX1 3PU, United Kingdom
}

\date{Accepted 2026 April 30. Received 2026 April 30; in original form 2026 March 16}

\pubyear{\the\year{}}

\begin{document}
\label{firstpage}
\pagerange{\pageref{firstpage}--\pageref{lastpage}}
\maketitle

\begin{abstract}
We examine the impact of progenitor age-dependent luminosity evolution of Type~Ia supernovae on a cosmographic measurement of the deceleration parameter $q_0$. 
Our recent redshift tomographic analysis showed that locally $q_0$ has a strong dipole anisotropy aligned approximately with the bulk flow, and only a small monopole component remains at distances exceeding a few hundred Mpc. 
Applying redshift-dependent corrections for progenitor age to the Pantheon+ catalogue, we find that this shifts the monopole component of $q_0$ to \emph{positive} values (i.e. deceleration), while leaving the local dipole component essentially unchanged.
\end{abstract}


\section{Introduction}
\label{sec:Intro}
It has been suggested that the cosmic acceleration inferred from Type Ia supernovae (SNe~Ia) could be illusory, due to our being `tilted' observers embedded in a bulk flow  \citep{Tsagas:2021tqa}. 
The inferred acceleration should then be directed mainly along the local bulk flow, and die out over a few hundred Mpc.
\cite{Colin:2019opb} analysed the SDSS-II/SNLS3 Joint Lightcurve Analysis (JLA) compilation of 740 Type Ia supernovae in the heliocentric frame (after removing erroneous corrections for pecular velocities) and found that the deceleration parameter $q_0$ is indeed mainly a dipole aligned approximately with the local bulk flow direction. 
This was criticised by \cite{Rubin:2019ywt} on the grounds that \cite{Colin:2019opb} ought to have first boosted to the CMB frame and applied corrections for peculiar velocities of the SNe~Ia host galaxies (`Hubble diagram' analysis). 
It has been demonstrated however that the distribution of matter as traced by quasars and radio galaxies is \emph{not} isotropic in the CMB frame \citep{Secrest:2020has,Secrest:2022uvx,Wagenveld:2023kvi,Bohme:2025nvu} --- for a review see \cite{Secrest:2025wyu}. 
This invalidates the criticism of \citet{Rubin:2019ywt} which rests on this assumption. 
Subsequently it was shown that for the Pantheon+ compilation \citep{Scolnic:2021amr} too, $q_0$ is mainly a dipole which dies out by $z \sim 0.1$ and is particularly pronounced in the frame of the Local Group \citep{Sah:2024csa}. 
Cosmic acceleration is thus likely a general relativistic effect due to the anomalous bulk flow in our local Universe, and not due to a Cosmological Constant $\Lambda$ --- as it must then be \emph{isotropic}.

The use of SNe~Ia as standard candles relies on the assumption that their  properties do not evolve with redshift. 
However \cite{Kang:2019azh} reported a $\sim3\sigma$ correlation between standardised SNe~Ia luminosity and the stellar-population age, based on spectroscopic observations of early-type host galaxies. 
The significance of the correlation was challenged by \citet{Rose:2020shp} but countered by \citet{Lee:2020usn}. 
\citet{Lee:2021txi} then clarified that it arises from a more fundamental dependence of the width-luminosity and colour-luminosity relations on the progenitor age, i.e at a fixed stretch/colour parameter SNe~Ia with younger progenitors are fainter. 

Recently, \cite{Chung:2025gsy} used an extended sample of 300 galaxies at redshift $z<0.45$ to estimate the host galaxy ages, finding a more significant correlation of up to $5.5\sigma$ with the Hubble residuals.
Building on this, \cite{Son:2025rdz} proposed a correction $\Delta m(z)$ for this redshift-dependent bias to the Phillips-Tripp formula for the SNe~Ia distance modulus (in e.g. the SALT2 light-curve template):
\begin{equation}
\mu_{\mathrm{SN}} = m_\text{B}-M +\alpha x_1 - \beta c - \Delta m (z)
\label{eq:age_bias_corr}
\end{equation}
where $m_\text{B}$ is the apparent magnitude (in
the rest frame ‘B’-band), $M$ the absolute magnitude, $x_1$ and $c$ are the `stretch' and `colour' corrections, and
\begin{equation}
\Delta m(z) = \Delta \text{age}(z) \times 0.030\,\text{mag}\,\text{Gyr}^{-1},
\label{eq:mag_corr}
\end{equation}
where $\Delta \text{age}$ is the change in mean progenitor-age of the SNe~Ia population (relative to $z = 0$) derived from the Supernova progenitor-age distribution (SPAD). This is obtained by convolving the SNe~Ia delay-time distribution with the cosmic star formation history. 
Individual SNe~Ia typically lack direct progenitor-age measurements, so the population-averaged $\Delta \mathrm{age}(z)$ provides an estimate of the average age evolution --- see Figs.~ 6 and 7 of \cite{Lee:2020usn}.
While the SPAD, and hence $\Delta \mathrm{age}(z)$, depend on the assumed cosmological model, \cite{Lee:2021txi} showed that this dependence is weak and does not significantly affect their main conclusions.
The value of the slope ($0.03\,\text{mag}\,\text{Gyr}^{-1}$) was taken as an average of measurements made by \cite{Chung:2025gsy} from the samples of \cite{Gupta:2011pa} and \cite{Rose:2019ncv}.\footnote{A new study by \cite{2026arXiv260113785W} criticises \cite{Chung:2025gsy} and \cite{Son:2025rdz} for not having done host-mass standardisation and argues that correcting for these removes any correlation between the Hubble residuals and the progenitor age. They also state that the progenitor age difference between nearby and distant supernovae was overestimated due to assumptions about the SNe~Ia delay time distribution. The debate continues.}
In this Letter, we test the impact of the progenitor-age-dependent luminosity evolution on the inferred cosmic acceleration and its direction dependence, in the statistically principled SNe~Ia analysis framework developed by us \citep{Sah:2024csa}.

\section{Method}
We briefly describe our methodology, referring to \cite{Sah:2024csa} for a detailed description.
The SNe~Ia luminosity distance, related to the distance modulus (\ref{eq:age_bias_corr}) as $\mu_{\mathrm{SN}} \equiv 25 + 5\text{log}_{10}(d_{L}/\text{Mpc})$, is Taylor expanded to 3rd-order in redshift, in terms of the Hubble rate $H_0 \equiv ({\dot{a}/a})|_{z=0}$, deceleration parameter $q_0 \equiv -({a\ddot{a}/\dot{a}^2})|_{z=0}$, and jerk $j_0 \equiv ({a^2\dddot{a}/\dot{a}^3})|_{z=0}$: 
\begin{equation}
\label{lum_dist}
\begin{aligned}
d_{L}(z) &= \frac{cz}{H_0}
\left[
1+\frac{1}{2}(1-q_0)z
- \frac{1}{6}\left(1 - q_0 - 3q_0^2 + j_0
+ \frac{kc^2}{H_0^2 a_0^2}\right)z^2
\right] \\
&\quad \times \frac{1+z_\text{hel}}{1+z}.
\end{aligned}
\end{equation}
Here $z_\text{hel}$ is the measured redshift in the heliocentric frame and $z$ is the cosmological redshift in the Cosmic Rest Frame --- usually identified with the CMB frame together with source peculiar velocity corrections --- which we denote as $z_\text{HD}$ (for `Hubble diagram'). 
We also show comparisons when it is just boosted to the CMB frame ($z_\text{CMB}$), as well as to the frame of the Local Group ($z_\text{LG}$). 
The multiplicative term accounts for our local peculiar velocity, ensuring that $d_L = (1 + z_\text{hel})\times \text{comoving distance}$ \citep{Rubin:2019ywt}. 
To ensure good convergence of the cosmographic expansion (\ref{lum_dist}) we impose a cut in redshift $z_\text{hel} \leq 0.8$ on the Pantheon+ catalogue, thereby excluding just 31 out of 1701 SNe~Ia (see \autoref{fig:example}).
We then adopt the Maximum Likelihood Estimator \citep{Nielsen:2015pga}: 
\begin{equation}
\begin{aligned}
\mathcal{L}(\theta)
    &= p\!\left[(\hat m_\text{B}, \hat x_1, \hat c)\mid\theta\right] \\
    &= \int p\!\left[(\hat m_\text{B}, \hat x_1, \hat c)\mid(M, x_1, c)\right]\,
       p\!\left[(M, x_1, c)\mid\theta\right]\,
       \mathrm{d}M\,\mathrm{d}x_1\,\mathrm{d}c ,
\end{aligned}
\label{eq:likelihoodC2}
\end{equation}
and the covariance matrix of \cite{Lane:2023ndt} --- as in analysis \textbf{C2} of \citet{Sah:2024csa} . Here the $~\hat{}~$ refers to the observed quantity and $p$ to the underlying probability
distribution of the true data. 
To reflect the expectation that an observer embedded in a bulk flow will infer a dipolar modulation of the deceleration parameter \citep{Tsagas:2015mua}, we model it as:
\begin{equation}
q_0 = q_\text{m} + \bm{q}_\text{d} \cdot\ \hat{n} \text{e}^{-z/S} 
\label{eq:scale-dependent}
\end{equation}
where $\hat{n}$ is the direction towards individual SNe~Ia, and the direction of the dipole ($\bm{q}_\text{d}$) is initially fixed to the CMB dipole direction. 

\subsection{Correcting for progenitor age bias}

\begin{figure}
 \includegraphics[width=0.9\columnwidth]{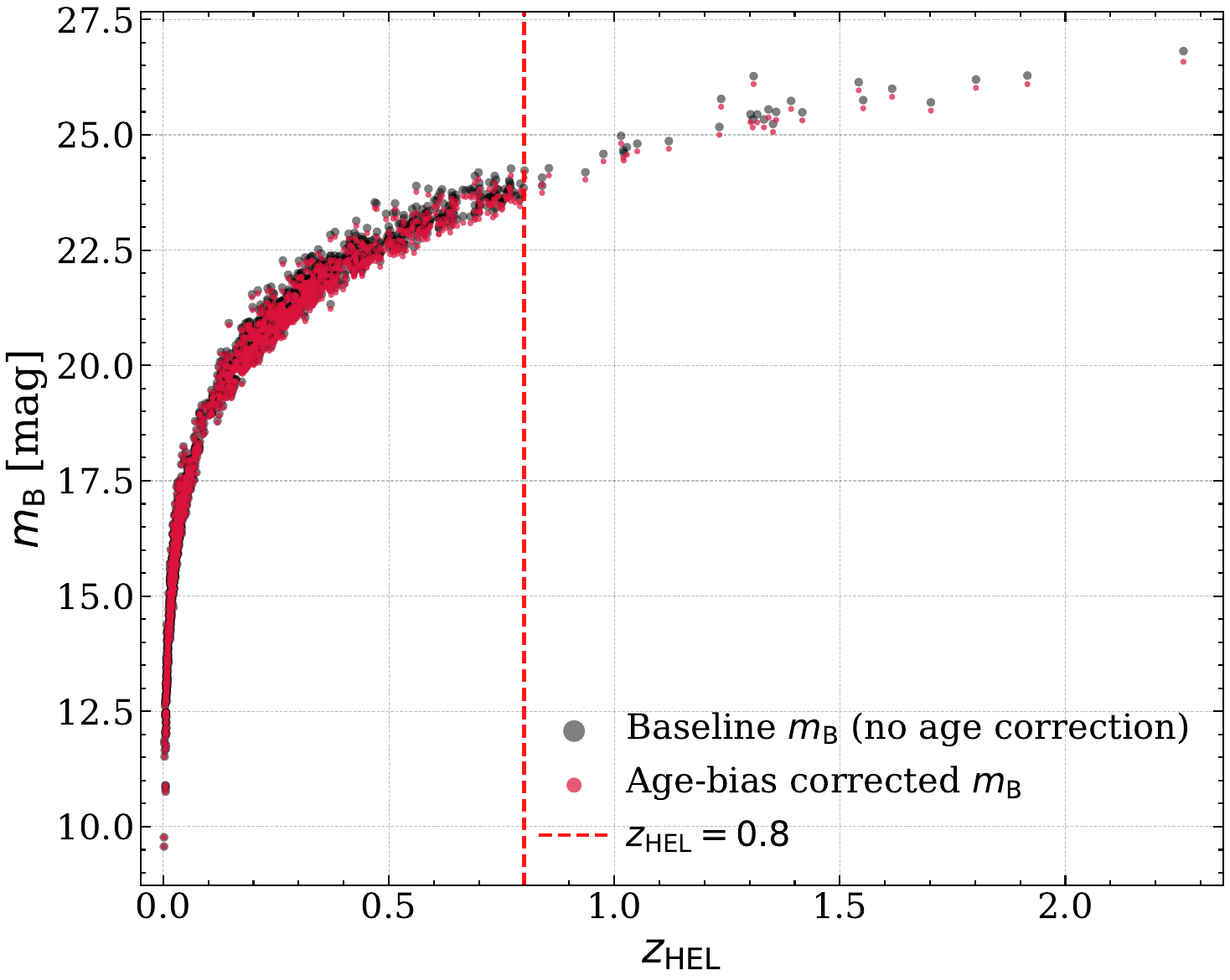}
 \caption{Apparent B-band magnitude as a function of heliocentric redshift for the Pantheon+ SNe~Ia catalogue. The black points show the baseline magnitudes ($m_\text{B}$) without progenitor age correction, while the red points ($m^{*}_\text{B}$) are corrected for progenitor age following \protect\cite{Son:2025rdz}.}
\label{fig:example}
\end{figure}

Following \cite{Son:2025rdz}, we apply the correction (\ref{eq:mag_corr}) to the apparent magnitudes of the Pantheon+ SNe~Ia (see \autoref{fig:example}):
\begin{equation}
    m^{*}_\text{B} =  m_\text{B} - \Delta \text{age}(z) \times 0.030 \text{ mag} \text{ Gyr}^{-1}.
    \label{eq:agebias_mb}
\end{equation}
Using 
the data for the redshift evolution of the median relative age in Fig.~2 of \cite{Son:2025rdz}, we use a cubic spline to interpolate the corresponding values of $\Delta \text{age}(z)$ at the desired redshift. 
In \cite{Son:2025rdz}, the redshift evolution of the stellar population age was derived using the $w_0-w_a$ CDM model fitted to DESI and CMB data. Since progenitor age estimates are mildly cosmology dependent, we  test later the sensitivity of our results to alternative  assumptions.

\section{Results}
\begin{figure*}
    \centering
\includegraphics[width=\textwidth]{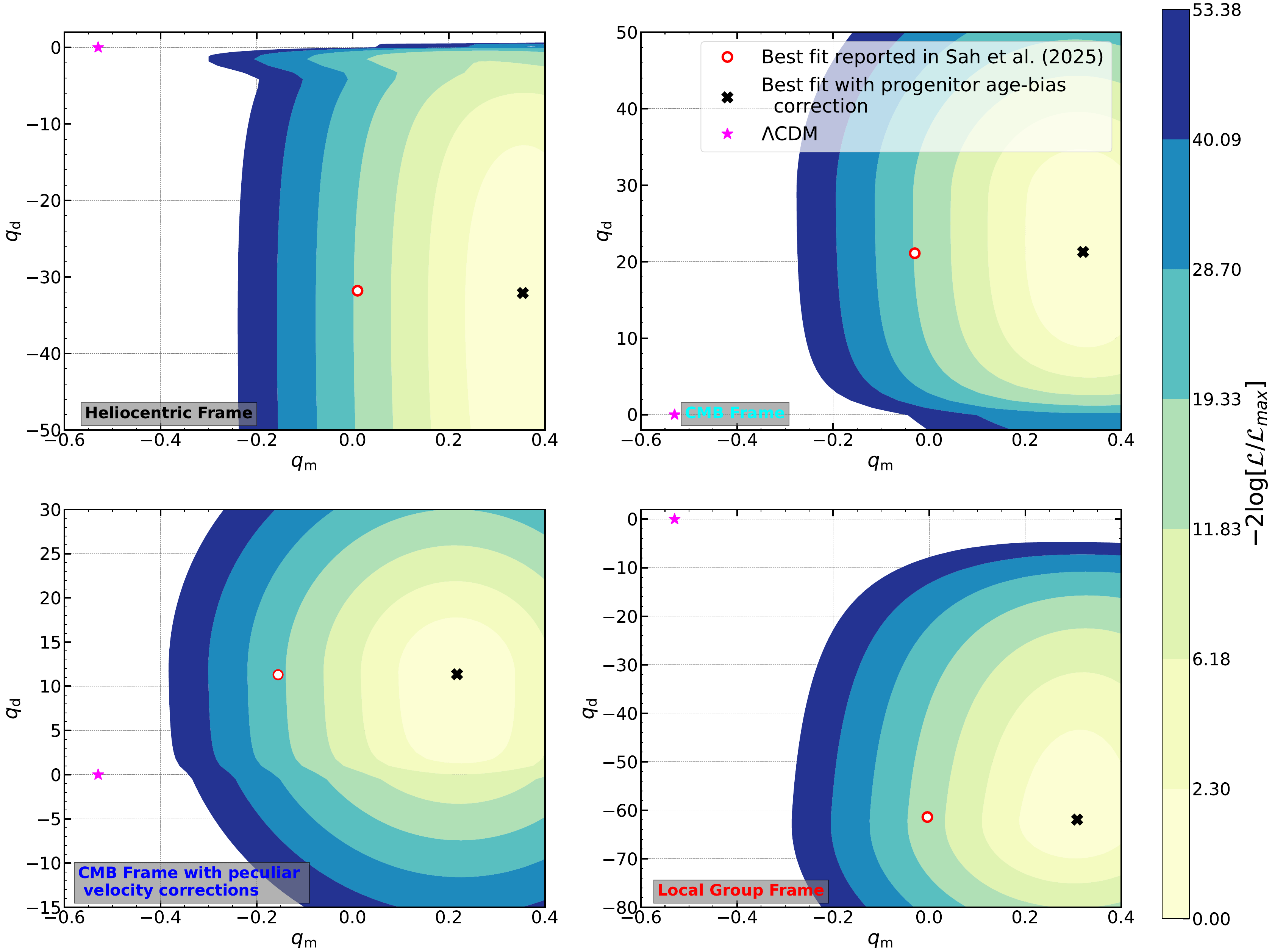}
    \caption{Contours at 1, 2, 3, 4, 5, 6 and $7\sigma$ for $q_\mathrm{m}$ and
    $\protect\bm{q}_\mathrm{d}$ in the heliocentric, CMB, Hubble diagram (CMB frame with peculiar velocity corrections) and Local Group frames, for Pantheon+ SNe~Ia with a redshift cut: $0.00937<z<0.8$. 
    The black cross is the best-fit with the progenitor age  correction, while the red circle is the  best-fit without such correction reported earlier \protect\citep{Sah:2024csa}. The magenta star denotes the expectation in the $\Lambda$CDM model \citep{ParticleDataGroup:2024cfk}.}
\label{fig:qm_qd_C2}
\end{figure*}

We employ the age-bias-corrected magnitudes (\ref{eq:agebias_mb}) in the likelihood (\ref{eq:likelihoodC2}), with a cut $z_\text{hel}>0.00937$ and show the results in \autoref{tab:qd_agebias_C2}, both with and without progenitor age bias corrections. 
While the statistical significance of the dipole remains unchanged, the monopole component shifts towards more positive values.
The dipole direction was fixed to the CMB dipole direction but allowing this to vary does not change our conclusions -- the best-fit dipole direction remains consistent within uncertainties,
Note that $\bm{q}_\mathrm{d}$ reverses sign in going from the heliocentric (or Local Group) to the CMB (or Hubble diagram) frames.
\footnote{
The redshift boosted to the Local Group (LG) frame is obtained from: 
$(1+z_\text{LG}) = (1+z_\text{hel}) \times (1+z_{\text{LG-hel}})$, where $z_{\text{LG-hel}}= \sqrt{(1-\vec{v}_{\text{LG-Sun}}\cdot\hat{n}/c)/(1+\vec{v}_{\text{LG-Sun}}\cdot\hat{n}/c)}-1$ and $\vec{v}_{\text{LG-Sun}}$ is the velocity of the LG frame relative to the heliocentric frame. 
The Sun's motion around the Galaxy is nearly in the \emph{opposite} direction to the CMB dipole hotspot, so while the heliocentric frame moves wrt the CMB at $369.82 \pm 0.11$~km\,s$^{-1}$ towards $l = 264.021^\circ \pm 0.011^\circ, b = 48.243^\circ \pm 0.005^\circ$ (which yields $z_\text{CMB}$ from $z_\text{hel}$), the LG moves wrt the CMB at $620 \pm 15$~km\,s$^{-1}$ towards $l = 271.9^\circ \pm 2.0^\circ, b = 29.6^\circ \pm 1.4^\circ$~\citep{Planck:2018nkj}.} 
Also, as found earlier \citep{Sah:2024csa}, the scale $S$ on which it decays (\ref{eq:scale-dependent}) corresponds to a redshift of $z\sim0.01$ i.e. $30h^{-1}$~Mpc, as expected if it indeed arises due to the local bulk flow \citep{Tsagas:2021tqa}.
\autoref{fig:qm_qd_C2} shows the $(1-7)\sigma$ contours around the best fit $q_m$--$\bm{q}_\mathrm{d}$ values listed in 
\autoref{tab:qd_agebias_C2}, constructed using Wilks’ theorem.  

\begin{figure*}
\centering
\includegraphics[width=\textwidth]{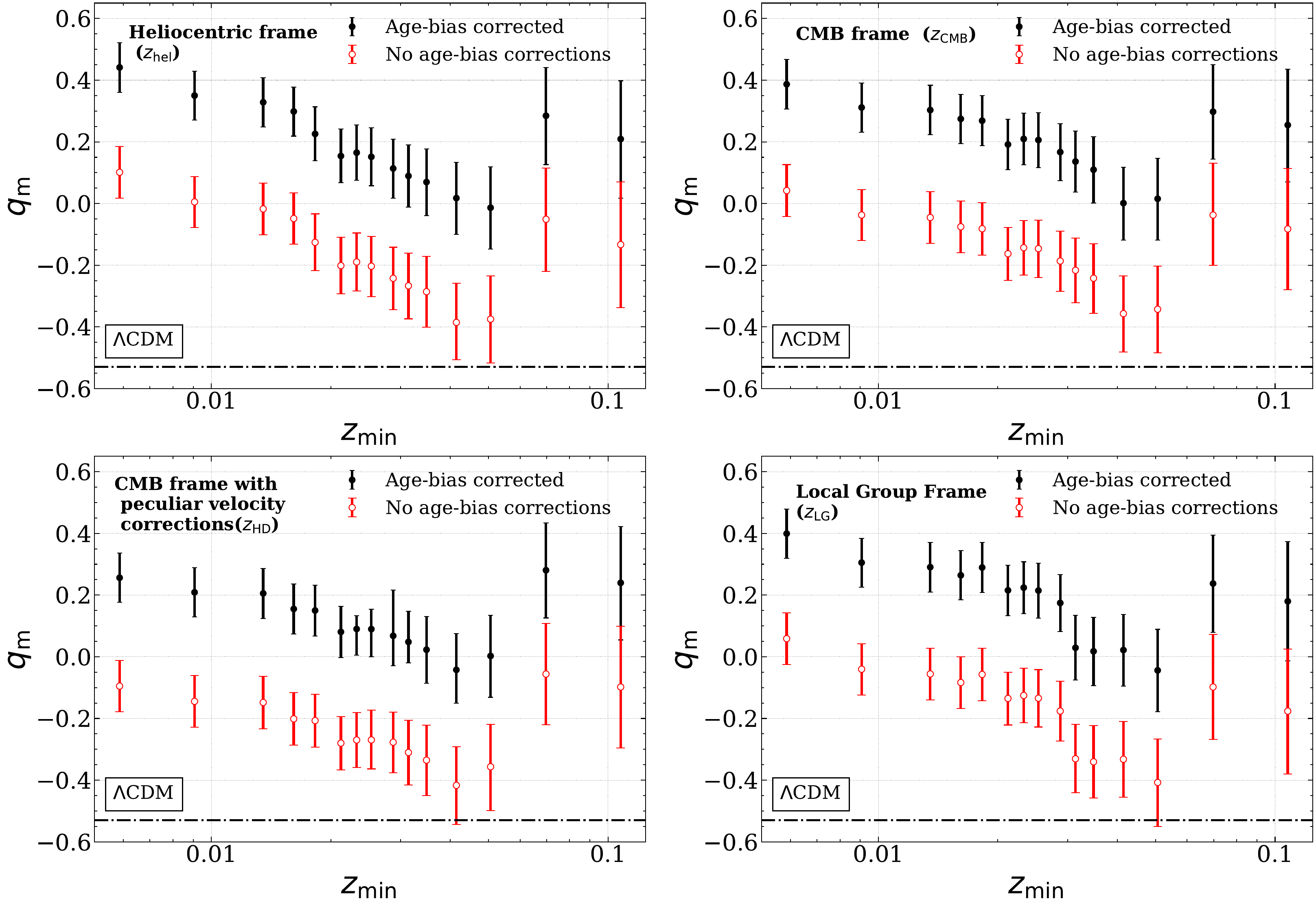}
    \caption{The monopole $q_\text{m}$ of the deceleration parameter for Pantheon+ SNe~Ia  with progressively higher cuts in redshift ($z>z_{\text{min}}$): in the heliocentric, CMB, Hubble diagram (CMB with peculiar velocity corrections) and Local Group frames. 
    Error bars indicate $1\sigma$ uncertainties obtained using Wilk's theorem. Black and red  points indicate the measurements with and without age-bias corrections. The dashed horizontal line indicates $q_\text{m} = -0.53$, the $\Lambda$CDM expectation.} 
\label{fig:qm_cum}
\end{figure*}

In \autoref{fig:qm_cum}, we plot $q_\text{m}$ moving the low-redshift supernovae in 50 SNe~Ia per step. 
It is evident that the age bias correction to SNe~Ia magnitudes removes any indication of accelerated expansion. 
This is consistent with the findings of \cite{Son:2025rdz} who too reported a currently decelerating expansion after accounting for progenitor age–dependent luminosity evolution.

\begin{table}
\centering
\setlength{\tabcolsep}{2pt}
\begin{tabular}{lcccccccccc}
\hline
 & \multicolumn{5}{c}{Baseline (no age corrections)} & \multicolumn{5}{c}{With age bias corrections} \\
Frame &
$\bm{q}_\mathrm{d}$ &$q_\mathrm{m}$ &$S$ &$\Delta\ln\mathcal{L}$ &$\alpha$ &$\bm{q}_\mathrm{d}$ &$q_\mathrm{m}$ &$S$ &$\Delta\ln\mathcal{L}$ &$\alpha$ \\
\hline
Hel &
$-31.8$ &$0.01$ &0.0094&$36.9$ &$5.7$ &$-32.1$ &$0.35$ &0.0094&$36.9$ &$5.7$ \\
LG &
$-61.4$ &$-0.04$  &0.0099&$135.5$ &$>7$ &$-62.0$ &$0.31$ &0.0099&$135.1$ &$>7$ \\
CMB &
$21.1$ &$-0.03$ &0.011&$31.8$ &$5.3$ &$21.3$ &$0.32$&0.011 &$31.4$ &$5.3$ \\
HD&
$11.4$ &$-0.14$ &0.01&$7.3$ &$2.2$ &$11.4$ &$0.21$ &0.01&$7.1$ &$2.2$ \\
\hline
\end{tabular}
\caption{
Best-fit dipole amplitude $\bm{q}_\mathrm{d}$, monopole $q_\mathrm{m}$,
likelihood ratio $\Delta\ln\mathcal{L}(q_\mathrm{d}=0)$, and statistical
significance $\alpha$ (in $\sigma$), before and after applying the progenitor age-bias correction. Results are shown in the heliocentric, Local Group, CMB, and Hubble diagram (CMB frame with peculiar velocity corrections) frames. 
The dipole amplitude (and the scale on which it decays) is unchanged by the correction, while the monopole shifts to positive values.
}
\label{tab:qd_agebias_C2}
\end{table}

\begin{figure}
\centering
\includegraphics[width=\columnwidth]{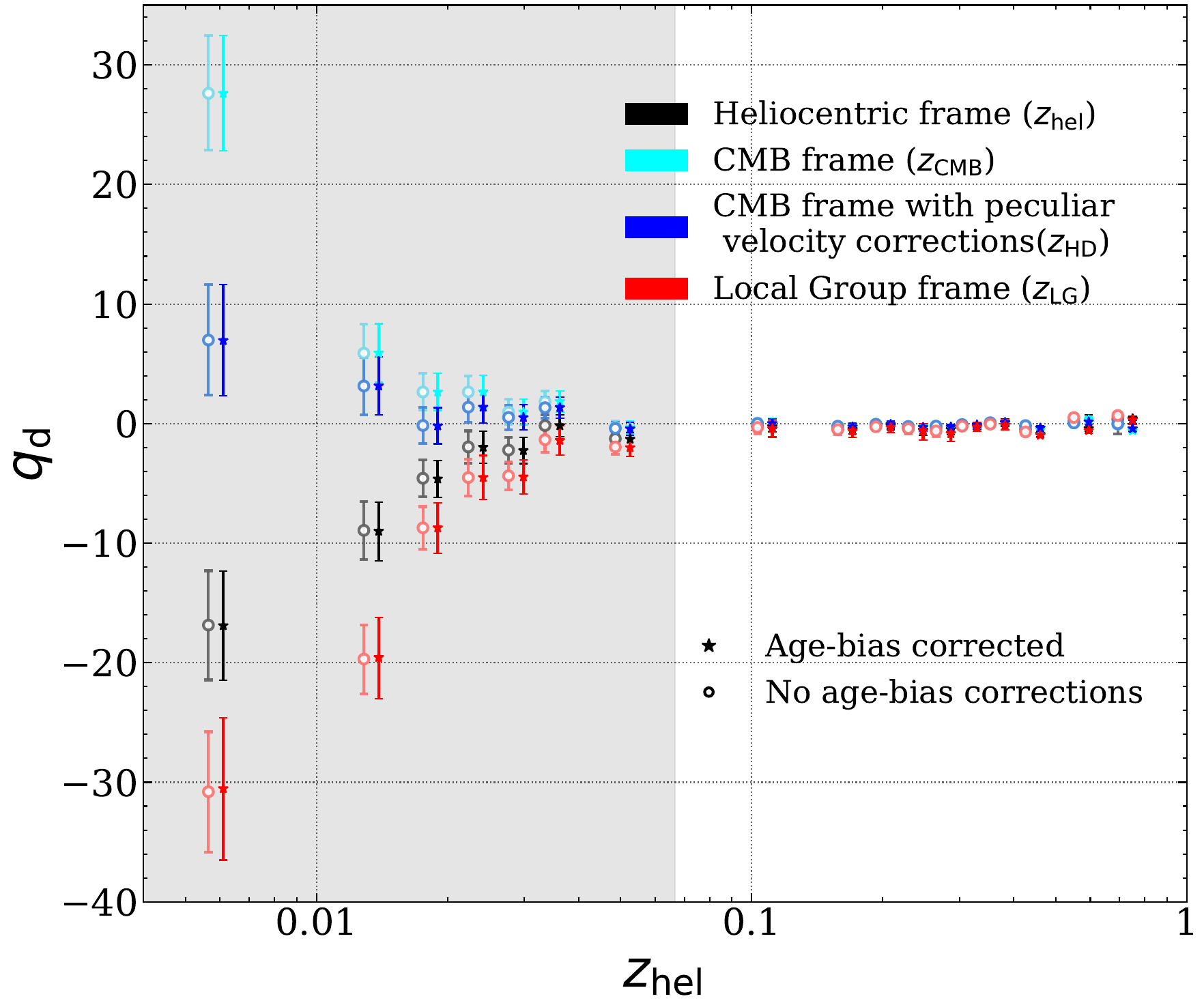}
    \caption{The local dipole $\bm{q}_\text{d}$ in the deceleration parameter evaluated in 17 shells containing around 100 SNe~Ia each, plotted against their median redshift. 
    The parameterisation is scale-independent, i.e. $q_0=q_\text{m} + \bm{q}_\text{d}\cdot\hat{n}$ within each shell. 
    Open circles and stars denote results without and with progenitor age–dependent luminosity evolution corrections. Analyses are performed in the heliocentric, CMB, Hubble diagram (CMB with peculiar velocity corrections) and Local Group frames, with the direction fixed to the CMB dipole. Error bars are 1$\sigma$ and age-corrected points are offset in $z$ for clarity. 
    The gray shaded region indicates $z<0.0667$ i.e. distance $<200h^{-1}$~Mpc. } 
\label{fig:shell}
\end{figure}

We also perform a scale-\emph{independent} fit to the dipole in $q_0$:
 \begin{equation}
q_0 = q_\text{m} + \bm{q}_\text{d} \cdot \hat{n},
\end{equation}
employing 17 shells, each containing 100 SNe~Ia (the highest redshift shell contains just 53), with the direction of the dipole  fixed to the CMB dipole direction. In \autoref{fig:shell}, we show the  dipole amplitude versus the median redshift of each shell. 
It is seen that the dipole amplitude $q_d$ remains unchanged, and its redshift dependence is consistent with that found earlier by \cite{Sah:2024csa} without any progenitor age bias corrections. 

As noted in \autoref{sec:Intro}, the age estimates for SNe~Ia progenitors are cosmology dependent. To assess this we repeat our analysis using the $\Delta$age values reported for two models by \cite{Lee:2021txi}: $\Omega_{\Lambda} = 0.73$, $\Omega_\text{m} = 0.27$ and $\Omega_{\Lambda} = 0$, $\Omega_\text{m} =0.27$. 
The differences are small and do not affect the inferred dipole amplitude or its statistical significance.

We also repeat the analysis with a cut $z_\text{hel} < 0.5$, finding no qualitative change in our results. 
As noted earlier \citep{Sah:2024csa} this is also the case when the entire Pantheon+ catalogue is included.
 
\section{Conclusions}
Although the evidence for a SNe~Ia progenitor-age bias \citep{Kang:2019azh,Lee:2020usn,Lee:2021txi} is still under discussion \cite{2026arXiv260113785W}, it is notable that correcting for these leaves the local dipole $q_\text{d}$ in the inferred deceleration parameter \citep{Sah:2024csa} unchanged within uncertainties. 
This correction rather turns the monopole component $q_\text{m}$ positive. 
There is thus no evidence for \emph{isotropic} accelerated expansion of the Universe, that can be ascribed to either a Cosmological Constant $\Lambda$ or more general dark energy.
\footnote{We note that in contrast to \citet{Sah:2024csa}, \citet{Zhou:2025rvf} report ``no hint of anisotropy'' in the Hubble expansion rate in their analysis of the Pantheon+ catalogue. 
They employ the redshifts $z_\text{HD}$ (CMB frame with source peculiar velocity corrections) using which the anisotropy in $H_0$ is minimised, as shown in Fig.~3 of \citet{Sah:2024csa} (and for $q_0$ in \autoref{fig:qm_qd_C2}). They also integrate over redshift in each direction, thus diluting the significance of the anisotropy, which is more prominent at low redshift as seen in Fig.~3 of \citet{Sah:2024csa} (and for $q_0$ in \autoref{fig:shell}). 
We have  argued elsewhere \citep{Rameez:2024xsn} that such corrections --- for our motion, as well as that of the SNe~Ia host galaxies, with respect to the CMB frame --- constitutes circular reasoning around an \emph{assumed} background FLRW cosmology. 
Moreover the recently established $>5\sigma$ mismatch between the dipole in the CMB and that in cosmologically distant quasars and radio sources \citep[see:][]{Secrest:2025bcw,Secrest:2025wyu} is a direct challenge to the FLRW assumption. 
Hence boosting to the CMB frame and applying peculiar velocity corrections to the sources (in order to emulate the Cosmic Rest Frame) is questionable.}

\section*{Acknowledgements}
We thank Young-Wook Lee for helpful communications.
This work was carried out while SS was visiting the Tata Institute of Fundamental Research, supported by a VAIBHAV Fellowship awarded by the Department of Science \& Technology, Government of India.

\section*{Data Availability}
 The code for reproducing our results is available at:
{\tt https://github.com/Shin107/Anisotropy-in-Pantheon-Plus}. 


\bibliographystyle{mnras}
\bibliography{Pantheon+AgeEvolution} 


\appendix

\section{Alternative analysis (\textbf{C1})}

To enable fair comparison with \citet{Rubin:2019ywt}, we also employ a second likelihood --- called \textbf{C1} in \citet{Sah:2024csa}:
\begin{equation}
\mathcal{L}[\theta]=\int p[\hat{m}_{B\text{corr}}|M] \times p[M|\theta] \text{d}M ,
\label{eq:likelihoodC1}
\end{equation}
This uses SNe~Ia magnitudes that have been corrected with sample- and redshift-\emph{dependent} stretch ($x_1$) and colour ($c$) corrections, as advocated by \cite{Rubin:2016iqe,Rubin:2019ywt}. These make the monopole of the deceleration parameter more negative as noted earlier  \citep{Rubin:2019ywt,Sah:2024csa}. 

The age-bias correction is now incorporated as:
\begin{equation}
    m_{B\text{corr}} = m_{B\text{corr}} - \Delta \text{age}(z) \times 0.030 \text{ mag} \text{ Gyr}^{-1}.
    \label{eq:agebias_mbcorr}
\end{equation} 
The results with and without age bias corrections are shown in \autoref{tab:qd_agebias_C1}.
The effect of progenitor age bias corrections is to shift $q_\text{m}$ to more positive values which are consistent  with zero, i.e. expansion at a constant rate.

\begin{table}
\centering
\setlength{\tabcolsep}{2pt}
\begin{tabular}{lcccccccccc}
\hline
 & \multicolumn{5}{c}{Baseline (no age corrections)} & \multicolumn{5}{c}{With age bias corrections} \\
Frame &
$\bm{q}_\mathrm{d}$ &$q_\mathrm{m}$ & $S$ &$\Delta\ln\mathcal{L}$ &$\alpha$ &$\bm{q}_\mathrm{d}$ &$q_\mathrm{m}$ & $S$ &$\Delta\ln\mathcal{L}$ &$\alpha$ \\
\hline

Hel &
$-6.27$ &$-0.37$ &0.024&$30.5$ &$5.2$ &$-6.27$ &$-0.01$ &0.024&$30.5$ &$5.2$ \\
LG &
$-39.6$ &$-0.41$ &0.012&$119.7$ &$>7$ &$-39.7$ &$-0.05$ &0.012&$119.3$ &$>7$ \\
CMB &
$20.5$ &$-0.39$ &0.011&$28.3$ &$5.0$ &$20.6$ &$-0.03$ &0.011&$27.9$ &$4.9$ \\
HD&
$11.4$ &$-0.49$ &0.01&$6.3$ &$2.0$ &$11.4$ &$-0.12$ &0.01&$6.2$ &$2.0$ \\
\hline
\end{tabular}
\caption{
    Best-fit dipole amplitude $\bm{q}_\mathrm{d}$, monopole $q_\mathrm{m}$, likelihood ratio $\Delta\ln\mathcal{L}(q_\mathrm{d}=0)$, and statisticalsignificance $\alpha$ for analysis \textbf{C1}, shown before and after applying the progenitor age-bias correction. 
    Results are presented in the heliocentric, CMB, Hubble diagram (CMB frame with peculiar velocity corrections) and Local Group frames. 
    The dipole in $q_0$ remains unchanged by the correction while its monopole component becomes close to zero.
}
\label{tab:qd_agebias_C1}
\end{table}

\bsp	
\label{lastpage}
\end{document}